# Cosmology at Low Redshifts

Ermanno F. Borra,
Département de Physique, Université Laval, Canada G1K 7P4
Email: borra@phy.ulaval.ca

It is argued that it is far more cost effective to carry out some projects with medium-sized dedicated zenith telescopes rather than large steerable telescopes, freeing the later to carry out projects that truly need them. I show that the large number of objects observed with a surveying 4-m zenith telescope allows one to carry out cosmological projects at low redshifts. Examining two case studies, I show first that a variability survey would obtain light curves for several thousands of type Ia supernovae per year up to z=1 and easily discriminate among competing cosmological models. Finally, I discuss a second case study, consisting of a spectrophotometric survey carried out with interference filters, showing its power to discriminate among cosmological models and to study the large-scale distribution of galaxies in the Universe.

## 1. Introduction

In most cosmological studies (e.g. searches for the cosmological parameters), the traditional strategy has been to observe at the highest possible redshifts with the largest telescopes since cosmological effects increase strongly with redshift. This strategy makes sense in a world where conventional large telescopes are expensive and few and must be shared among many observing projects. Unfortunately, telescope time is assigned by committee on a competitive basis so that even a high-priority project can only obtain a handful of nights per year. Consequently, the small number of observations that one can obtain gives large statistical error bars so that to optimize the signal-to-noise ratio one observes where the effect is large, namely at high redshifts. However, this strategy suffers from severe problems. A major well-known difficulty comes from the fact that evolutionary effects increase strongly with redshift. This makes it difficult to use high redshift objects for cosmological purposes, for evolutionary and cosmological effects are intertwined and difficult to disentangle. A second drawback comes from the small numbers of objects observed. This not only gives high statistical errors but there is a far more worrisome and insidious penalty: it makes it difficult to understand the data and identify systematic errors.

In this article, I show that it is a far better strategy to observe at relatively low redshifts, where the cosmological effects are smaller, with a medium-size dedicated zenith telescope. The reason is that one can gather massive quantities of observations that give very small statistical errors (from Poisson statistics) and, most importantly, allow one to understand systematic errors. I consider two specific examples of the power of large surveys to discriminate between cosmological models at intermediate redshifts. I shall assume a dedicated zenith 4-m survey telescope. I use cost as a primary design driver, an unusual consideration in an Astronomy article.

**2. Observing with a 4-m Dedicated Zenith Telescope**

Because cosmological objects are faint and one has to observe many objects, it will be difficult to carry out cosmological observations with an instrument much smaller than a 4-m class telescope. To observe a number of object sufficiently large that one can detect the small cosmological effects predicted at low z (see sections 3 and 4 in this article) one needs a 4-m telescope working full-time on the project. Because fully steerable 4-m class telescopes are expensive, it will be difficult to obtain the funds to build such an instrument and dedicate it full time to a single project. Fortunately, as I show below, it is not necessary to use a fully steerable telescope since one can do it with a fixed zenith pointing telescope. A fixed zenith telescope will be far less expensive, as argued in the discussion.

A zenith telescope that tracks with a CCD in the Time Delayed Integration mode (also referred to as driftscan) observes with an integration time given by the time it takes an object to drift across the CCD detector. The nightly single-pass time (in seconds) is given by

$$t = 1.37\ 10^{-2}\ n\ w / (f \cos(lat)), \qquad (1)$$

where n is the number of pixels along the read-out direction of the CCD, w the pixel width (microns), f the focal length of the telescope (meters) and *lat* is the latitude of the observatory. TDI tracking with a zenith telescope aberates the PSF but there are correctors that can remove most of it (Hickson & Richardson 1998).

A 4-m diameter f/2 telescope equipped with a 4KX4K CCD having 15-micron pixels ( 0.4 arcsec/pixel) obtains an integration time of 120-seconds/nightly pass. Table 1 gives the limiting magnitudes in B, R and I estimated with the IRAF task CCDTIME available on the NOAO web site, assuming the CCD prime focus camera at the Kitt Peak 4-m telescope. The zenith observations assume 1 arcsecond seeing and a 7-day-old moon.

Figure 1 converts equatorial into galactic coordinates and can be used to show the strip of sky sampled by a zenith telescope at a given site. The strip of sky follows a straight line of constant declination equal to the terrestrial latitude of the observatory. At a terrestrial latitude of + 30 degrees, it rises from the galactic plane, climbs through the North Galactic Pole and falls back to the plane; at -30 degrees it goes through the bulge of our galaxy and the South Galactic Pole.

The width of the strip of sky observed by the CCD is given by

$$S = 0.206\, n\, w / f, \qquad (2)$$

where S is expressed in arcseconds, w is the pixel width in microns, and n is the number of pixels in the direction perpendicular to the scan. With a 4KX4K CCD mosaic having 0.4 arcseconds pixels, the strip of sky observed is 26 arcminutes wide. Using spherical trigonometry on the celestial sphere, one finds that it observes 146 square degrees, 72 square degrees of which are “extragalactic” regions of low galactic obscuration having galactic latitudes > 30 degrees.

One may also consider carrying out, with a 4-m class zenith telescope, a spectrophotometric survey using narrow-band filters such as the planned LZT survey (Hickson et al 1998). The telescope will cover the wavelength region from 4,000 Å to 10,000 Å with 40 interference filters having logarithmically increasing widths and adequate overlap (Hickson et al 1998). Assuming a complete spectral coverage from 4000 Å to 10,000 Å, taking account losses to bad weather (in a good site), one would get 3 passes/filter/year for a total integration time of 360 seconds/filter. In 4 years one would get 12 passes for a total integration time of 1,440 seconds/filter. In four years, the performance of the telescope observing with a S/N = 10 per filter (with the exception of the reddest filters) can readily be estimated to reach B=23.3 and R = 23.1, sufficient to get 6,000 Km/sec to 1,000 Km/sec redshift errors, depending on the Hubble type (Cabanac and Borra 1995), and B=24 with S/N =5, sufficient for rougher energy distributions and redshifts. The increase in sky brightness with wavelength is roughly compensated by the flux increase with wavelength for most faint galaxies, at least for $\lambda <$ 7,000 Å.

**3. Case Study 1: a supernova strip survey**

A zenith telescope observes the same region of sky night after night and measures the fluxes of all objects in it every clear night. Faint objects can be observed in the visible region of the spectrum for two weeks of moonless nights a month while there is no restriction for observations redder than R. The sampling time is well matched to the time scales of SN light curves.

Table 2 gives the apparent magnitudes at maximum light of Type Ia supernovae as a function of redshift (Schmidt et al. 1998). Comparison of Tables 1 and 2 shows that a single nightly pass can detect a SNIa near maximum light to z=0.8. Furthermore, considering the 1+z time dilation factor, one can bin 6 nights at z>0.5, going 1 magnitude deeper, and to z > 1. Binning 6 nights in I is realistic if one considers that the I limiting magnitudes are essentially independent of the phase of the moon. Light decay time corrections to canonical peak luminosities need observations to 15 days in the rest frame, at which time the luminosity has decreased about 2 magnitudes below the peak, allowing one to observe useful light curves to $z = 1.0$.

SN rates are uncertain. Pain et al. (1996) estimate the rate of Type Ia supernovae at $z \sim 0.4$. They predict 34.4 (+23.9, -16.2) events/year/square degree for magnitudes in the range $21.3 < R < 22.3$, corresponding to $0.3<z<0.5$. Section 2 shows that a survey with a 4KX4K CCD covers an extragalactic strip of sky ( bII > 30 degrees) having a surface of 72 square degrees so that there should be about 3000 events per year in the strip. Approximately 2/3 of the objects would be unobservable in a given night because they would lie outside of the nightly strip monitored, reducing the number to about 1000 SNe Ia per year having $0.3<z<0.5$. Table 3 estimates the number of SNe Ia in other redshift ranges by extrapolating with the differential volume element dV(z, H0, q0)/dz . I assume H0 = 75 km/sec/Mpc, q0 = ½ and neglect evolution, an assumption likely to underestimate the counts for $z>0.5$. The numbers broadly agree with those in Wang (2000). The uncertainties in the observed rates do not warrant more sophisticated modeling. The uncertainties in these estimates are obvious and will not be discussed further.

As a bonus, the survey will give an unprecedented sample of variable stellar and extragalactic objects. There would be a vast variability database from which one could extract microlensed objects, extragalactic gravitational lenses, serependitous objects, etc…

Branch and Tammann (1992) have reviewed the use of type Ia supernovae as standard candles and consider several cosmological research topics. There is now intense interest in the determination of the cosmological parameters with Type Ia SNe. It took several years for the two major teams involved to gather data on a few 100 SNe. The 4-m survey would get several thousand multicolor light curves in a few years, allowing one to study and understand characteristics of the sample such as extinction and light-curve dependent luminosity effects. Other interesting applications would come from the several thousands calibrated radial rods so gathered. For example, this would allow one to study the velocity field and peculiar motions, mapping the mass distribution of the Universe. Finally, the rates of supernovae, of all types, can be used to infer the rate of massive star formation as function of z.

As a specific example, let us consider the power of the survey to discriminate among cosmological models. Figure 2 gives the magnitude-redshift relations for three "flat" cosmological models with different combinations of $\Omega_M$ and $\Omega_\Lambda$. For clarity, I have subtracted from all curves the magnitude-redshift relation for the $\Omega_M = 0.3$, $\Omega_\Lambda= 0.0$ model. The small vertical bars at z=0.2, z=0.4, z= 0.6, z=0.8 and z = =1 give the estimated 2 standard deviations error bars. To compute the error bars, I have assumed a 0.4 magnitude standard deviation for an individual SN, 1000 SN / year/(0.2 z bin) and that the magnitude errors are normally distributed, obtaining thus a standard deviation of 0.0125 magnitude for one year-worth of SN observations. We can see that the data easily discriminate among the models. If a systematic effect, such as dust or evolution, mimics a particular signature, it would have to mimic it very closely indeed. Podariu, Nugent, & Ratra (2001) have computed cosmological models with time-varying cosmological constants. Their figure 1 shows that most of the differences among models occur for $0.2<z<1.0$.

A bottleneck may arise from the need to do follow-up spectroscopy to confirm that one has a Type Ia SN with a conventional telescope. On the other hand, detailed multicolor light curves minimize, if not totally eliminate, the need for spectroscopic observations of the supernovae themselves. Redshifts are needed but, because the host galaxies will be closely packed inside a 1-degree wide strip, their redshifts can be efficiently obtained with multiobject spectroscopy. In the most pessimistic scenario, should follow-up spectroscopy of the SNe be necessary, the use of a zenith telescope to find and follow supernovae would save precious discovery time on oversubscribed conventional telescopes to be reallocated for spectroscopic follow-up. Note also that redshifts of the host galaxies can be obtained with the same 4-m telescope and interference filters (section 2 above and section 4 below).

**4. Case Study 2: $q_0$ from galaxy counts**

Let us consider the determination of $q_0$ from galaxy counts. This is a notoriously difficult measurement since curvature is small at low redshifts and one therefore traditionally has tried to obtain it, with a variety of methods, from observations at high redshifts. A first difficulty arises because, having to observe far, one needs intrinsically bright (or large) objects that tend to be rare. A second difficulty is caused by evolution effects, important given the great lookback time at high z, that have bedeviled efforts to get q0: geometry and evolution affect all tests and cannot be disentangled.

Volume tests give the most sensitive measurements for $q_0$. Consider for example, the number of objects /unit surface/unit redshift. It is given by

$$\frac{dN}{dz} = d\Omega N_0 c^3 [q_0 z + (q_0 - 1)(\sqrt{1+2q_0 z} - 1]^2 / [H_0^3 (1+z)^3 q_0^4 \sqrt{1+2q_0 z}], \qquad (3)$$

where the symbols have the usual meaning and $N_0$ is the space density at z=0.

As with all geometrical tests, the difference among the various geometries only becomes large for z >1, where only intrinsically bright rare objects are detectable and where evolution effects are large; but is small for z<0.3 where evolution is much smaller and where intrinsically fainter and more numerous objects (e.g. ordinary galaxies) are detectable. The difficulty is illustrated by the 3 curves in Figure 3 that show the relative counts predicted at z = 0.1, 0.3 and 0.5 as function of $q_0$, normalized to the counts predicted for q0=0.5. The differences among the curves are small and it is difficult to discriminate among them. This is why, traditionally, one has observed at high redshifts.

Let us determine whether we live in an open or closed Universe and take the criterion that we can differentiate between $q_0 = 0.4$ and $q_0 = 0.5$ at the 5 σ level. Figure 3 shows that at z = 0.3, the ratio between the counts for $q_0 = 0.4$ and for $q_0 = 0.5$ is 1.05. If we ask for a 5 standard deviation discrimination and assuming that $H_0$ and $N_0$ are known, Poisson statistics demand a minimum of 10,000 objects. Poisson statistics cannot be blindly applied because galaxies are known to be clustered on a scale of a few Mpc; there will be an excess of variance in the cell counts compared to a random distribution. However, the correction factor that takes into account the departure from Poisson statistics scales as 1/volume so that the correction is negligible for the large volume of sky that we sample, provided galaxies remain clustered at z=0.3 as they are in the local universe. This can be checked with the data.

To have an approximate view of how the observations would sample the Universe, I have computed the redshift distribution expected from a survey having lower limiting magnitude $m_0$ and upper limiting magnitude $m_1$ from the usual cosmological integral

$$\frac{dN}{dz} = d\Omega \int_{m0}^{m1} \Phi(M) \frac{dV}{dz} dm \quad , \tag{4}$$

where Φ(M) is the differential luminosity function (per unit magnitude) of galaxies, $M(H_0, q_0, z, m)$ the absolute magnitude, m the apparent magnitude, dΩ the surface area element and $dV(H_0, q_0, z)$ the cosmological volume element , $H_0$ the present epoch Hubble constant, $q_o$ the deceleration parameter and z the redshift. The models have been computed with the mix of galaxy types and the K-corrections described in Shanks, Stevenson, Fong, & MacGillivray (1984). This is an oversimplification for the luminosity function varies with morphology and the Shechter function is an average over all Hubble types. Furthermore, the luminosity function evolves in time and it is far from trivial to take evolution into account. I have approximately taken evolution into account by simply using the parameters of the luminosity function, appropriate for the redshift ranges and magnitudes involved, from Lilly et al. (1995). The uncertainties brought by these assumptions are large but tolerable for our purpose since we are only interested in an estimate of the redshift space sampled, rather than detailed modeling. Figure 4 shows the redshift distributions expected for surveys reaching 22nd, 24th and 27th blue magnitudes. Hudon & Lilly (1996) have also computed redshift distributions in the R band. It must be noted (e.g. Hudon & Lilly, 1996) that models underestimate galaxy counts at low apparent magnitudes for galactic evolution introduces an excess of high-redshift galaxies. By the same token, one must realize that the Universe will be sampled to higher redshifts than indicated in Fig. 4. Taking accounts the uncertainties of the models, Fig. 4 predicts about 1000 galaxies per square degrees with $0.35 < z < 0.45$. As discussed in section 2, the proposed telescope, located at a latitude of 30 degrees and equipped with a single 4KX4K CCD observes 74 square degrees of extragalactic sky. Therefore about 100,000 galaxies in the entire survey would have usable energy distributions. An independent check of these predictions can be made by considering Figure 3 in Hudon and Lilly (1996). It shows that there are about 3500 objects in the 4 square degrees CFRS survey with $0.35 < z < 0.45$ and $19 < R < 23.5$, essentially the magnitude range covered by the proposed survey. Their simulations predict that the telescope would observe about 50,000 galaxies with $0.35 < z < 0.45$, in reasonable agreement with the number here computed.

About 100,000 galaxies could thus be observed in the one-degree-wide strip accessible with a conventional corrector. It will not be possible to obtain redshifts for all objects and there will also be some loss due to some large nearby galaxies and the halos of bright stars. However, there should be a sufficient number of ordinary ellipticals and spirals to get the few 10,000 redshifts needed for a discrimination at the 5 σ level. Note also that over 1/3 of the galaxies should be ellipticals and early type spirals for which σ ~ 1000 Km/sec can be obtained from the 4,000 Å break (Cabanac & Borra 1995).

In practice, one would have to obtain counts at various redshifts and use a least-squares fitting procedure and he would have to consider systematic effects that may cause spurious z-dependent gains or losses of objects (e.g. magnitude cutoff, redshift or photometric errors). The large quantity of data should leave us well-equipped to understand this. For example, there is evidence that the faint end of the luminosity function evolves at surprisingly low redshifts. The data would allow us to measure the evolution and either correct for it or simply truncate the luminosity function at the appropriate magnitude. Close attention shall have to be paid to effects peculiar to the data, such as the effect of the large redshift errors, which depend on the Hubble type. This is not the place to carry out a detailed discussion; the relevant point is that the data allows us to contemplate such a project at all. The theme of this paper is that having a zenith telescope dedicated to a project allows us to consider a project that would be otherwise unthinkable.

## 5. Case study 3: Large Scale structure

Figure 4 shows that the universe is sampled to significantly higher redshifts than any other large scale existing redshift (e.g. the Sloan Digital Survey), albeit with a considerably lower radial velocity precision. One could therefore study the large scale structure of the universe from a few times the radial velocity precision to the redshift depth of the survey (a few Gpc). This complements the information that will be obtained from the other more precise but shallower surveys. Figure 5 (adapted from Vogeley 1995) shows the uncertainty of the SDSS power spectrum. The 1 σ uncertainty expected for a volume limited (to M*) sample of the SDSS northern redshift survey, assuming Gaussian fluctuations and a Ωh=0.3 CDM model, is compared to power spectra for CDM with different Ωh. Power bars on smaller scales are of similar or smaller size than the symbols. The shaded box shows the range of the HPBW of the z distribution of the spectroscopic survey (see Fig. 4). The HPBW of the spectroscopic survey extends well beyond the HPBW sampled by the SDSS. Because the total numbers of galaxies are similar in the 2 surveys, we can expect a similar distribution of error bars shifted to the shaded box. The error bars in the shaded box will approximately have the sizes of the error bars of the SDSS for λ<200 MPc. We can see that the spectroscopic survey samples, with small error bars, the very long wavelengths at which the differentations among the theoretical models is the greatest, and overlaps with the scales probed by COBE. Hopefully, the small error bars given by the large statistics may be able to detect the features predicted by some models; if needed the statistics can be increased by observing different strips of sky. The telescope could be moved or other ones build, an acceptable alternative given the low cost of the system.

Because we get energy distributions, morphologies and accurate photometry we can repeat the analysis as function of Hubble type, spectral type, etc.... We also can study the redshift dependences of the energy distributions and mixes of Hubble types.

## 6. Discussion and conclusion

The main message of this article is that it is far more cost effective to carry out some projects with medium-sized dedicated zenith telescopes rather than large steerable telescopes, freeing the later to carry out projects that truly need them. There certainly are many examples of astronomical observing programs presently carried out with large steerable telescopes that could be redesigned to be executed on zenith telescopes in order to reach the same scientific goals.

As a specific example, I argue that many cosmological studies could be carried out at low redshift with medium-sized telescopes, rather than large telescopes observing at high redshifts. Observing at low redshifts has several advantages, the main ones being that evolution effects are small and easier to quantify and that the large number of objects observed gives a better understanding of the data.  I argue that the only way this can be done is by using very low cost instrumentation. Not only would it be unlikely that one can find the money for a dedicated conventional telescope; it is not even necessary since a far less expensive zenith telescope can do the job.

Two case studies, a supernova survey and a galaxy survey illustrate the point quantitatively. I design the experimental execution of  the cosmological science drivers along instrumentation that minimizes costs. Building an observing program along financial considerations is something unusual in a scientific paper; but we must face the fact that Astronomers live in a world where costs matter and money is an issue one has to deal with. If one can afford it, it gets done; if one cannot afford it does not  get done: It is that simple. Consequently, to save costs, I assume that one will use a zenith telescope. With a zenith telescope savings are everywhere. There is no  moving frame; there is only a simple tower to hold the upper-end. There is no moving dome; there is only a simple silo-like enclosure with a roll-off roof. Because gravity always pull in the same direction, the upper-end is also much simpler and inexpensive than it is for a steerable telescope, since commercially available optical benches and mounts can be used.  The instrumentation looks more like laboratory instrumentation than conventional astronomical instrumentation that is expensive because it must be rugged. For example, one can use a simpler cold-box design and a commercial cooling unit positioned a few meters from the detector. Note also that the design discussed here uses a relatively small field of view; consequently the corrector is also small and relatively inexpensive (about $100,000).

Once one accepts that zenith telescopes are cost and science competitive instruments, he comes to the realization that the cost of the observatory is dominated by the cost of the mirror. Another major cost saving then becomes obvious: use a liquid mirror instead of a glass mirror. Liquid mirror telescopes have come of age for the technology is well-proven both in the laboratory, where it has demonstrated high optical qualities and robustness (Girard & Borra 1997, Tremblay & Borra 2000), and in observatory settings (Cabanac & Borra 1998, Hickson & Mulrooney 1997) where it has demonstrated scientific results and long-term reliable performance. The low cost advantage of liquid mirror telescopes allows one to dedicate one to a narrowly defined project that would be unpractical with a classical telescopes. As a rule of thumb, a liquid mirror and its supporting hardware cost two orders of magnitude less than a glass mirror having the same size and quality. The low capital and operating costs of liquid mirrors and liquid mirror observatories are well-documented (Tremblay 1999, Mulrooney 2000). These estimates are particularly robust because the NODO telescope has been run by NASA for several years (Mulrooney 2000). A zenith telescope can easily be robotized or remote controlled, therefore further reducing costs by eliminating the need for a night operator. Table 4 estimates the cost of building an observatory that houses a fully instrumented 4-m telescope. Appendix I elaborates on how this estimate was reached. Table 2 in Hickson et al. (1994) gives the cost breakdown of an observatory that housed a 2.7-m LMT. The costs in Table 4 compared to those of an observatory that was constructed and operated seem reasonable, considering the difference in size. Table 5 estimates the costs of operating the observatory. Appendix I elaborates on how this estimate was reached..

I have, on purpose, restrained the case studies to topics of current interest and telescope sizes that are relatively small. However, it is clear that future observing programs and instrumentation would have to change to reflect changing science drivers. For example, a single 4-meter zenith telescope could give place to arrays of larger zenith telescopes. The design of the instrument could be radically different. A good example is the LAMA telescope, that will use an array of liquid mirrors (Hickson & Lanzetta 2002). Given the science drivers, the challenge to using zenith telescopes consists in design the observing program so as to maximize their advantages and minimize their disadvantages.

Carrying out a research project with a zenith telescope will free precious observing time on large steerable telescopes, time that can then be dedicated to the large quantity of other projects that truly need a large collecting area.

Serendipity may turn out to be the most important discovery of a large dedicated survey telescope. The history of Astronomy shows that whenever a new frontier (e.g. radio waves, γ-rays) has been opened, Astronomy has made huge strides, often led by unexpected discoveries (quasars, γ-ray bursts). Dedicated telescopes open a new frontier: The statistical frontier of extremely large numbers.

## 7. Acknowledgments

This work was supported by the Natural Sciences and Engineering Research Council of Canada. I wish to thank Drs. P. Hickson, M.K. Mulrooney and E.H.

Richardson for discussions that helped to determine the capital and operating costs of the 4-m observatory.

**Appendix I: Estimates of the Costs of a 4-m Zenith Telescope.**

Figure AI1 shows the schematics of a zenith telescope. The top end consists of a focusing system and a detector: there is some cost saving in the upper end structure since it does not have to be tilted. A large cost savings obviously come from the frame which consists of a simple tripod.

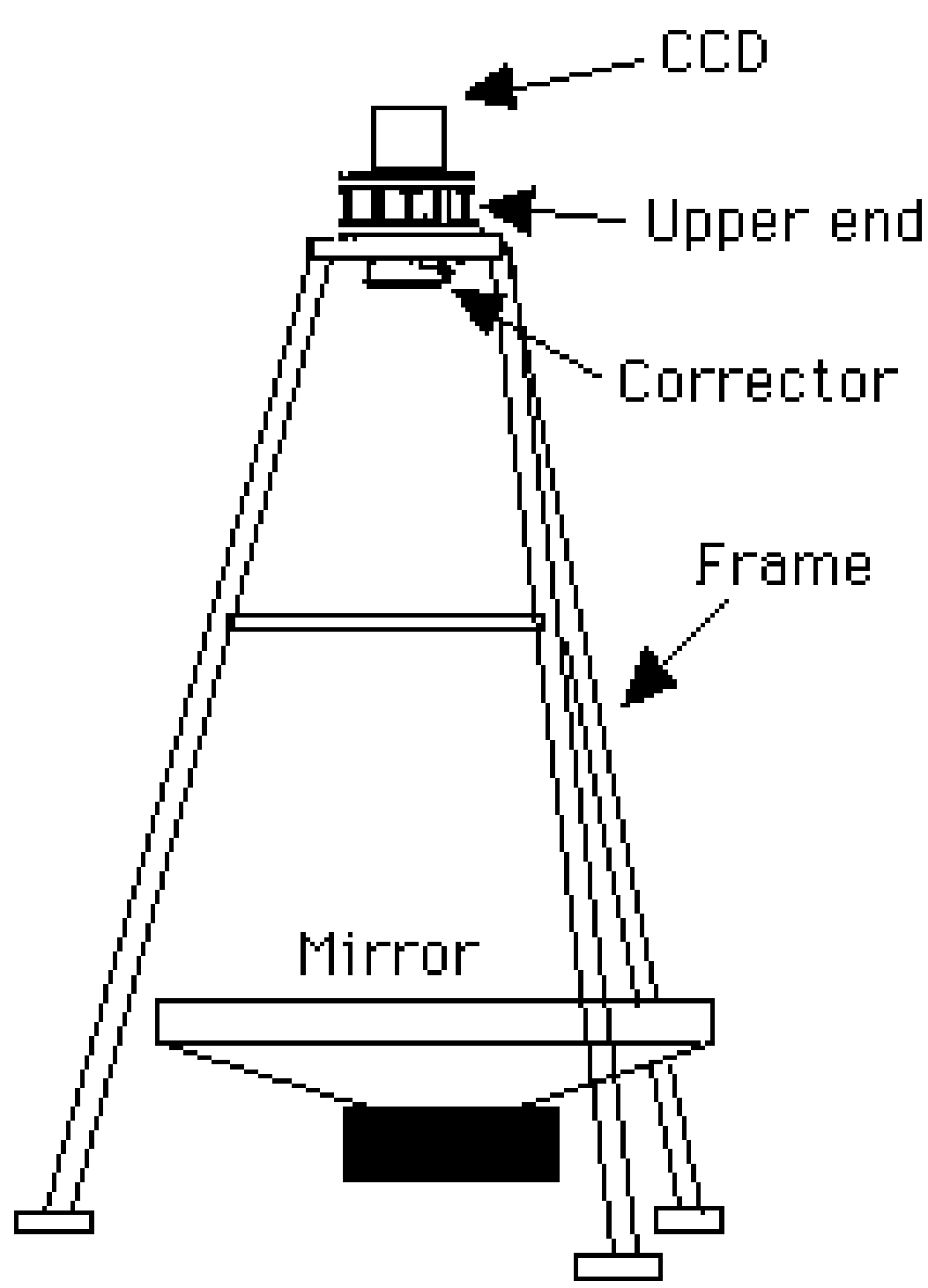

Figure AI.1: It shows the entire telescope system.

Figure AI.2 shows a layout of the telescope and observatory for a 4-m telescope. The structure is much simpler than the dome of a conventional telescope. The roof and the folding platform needed to service the upper end are the only movable parts and are inexpensive. The structure consists of a steel frame with metal sidings. Table 4 gives a cost estimate breakdown of the entire system. It does not give an estimate for the cost of the detector since it depends on the type of detector and, furthermore, prices vary with time. The compressor shack houses a compressor, needed if one uses a liquid mirror with an air bearing.

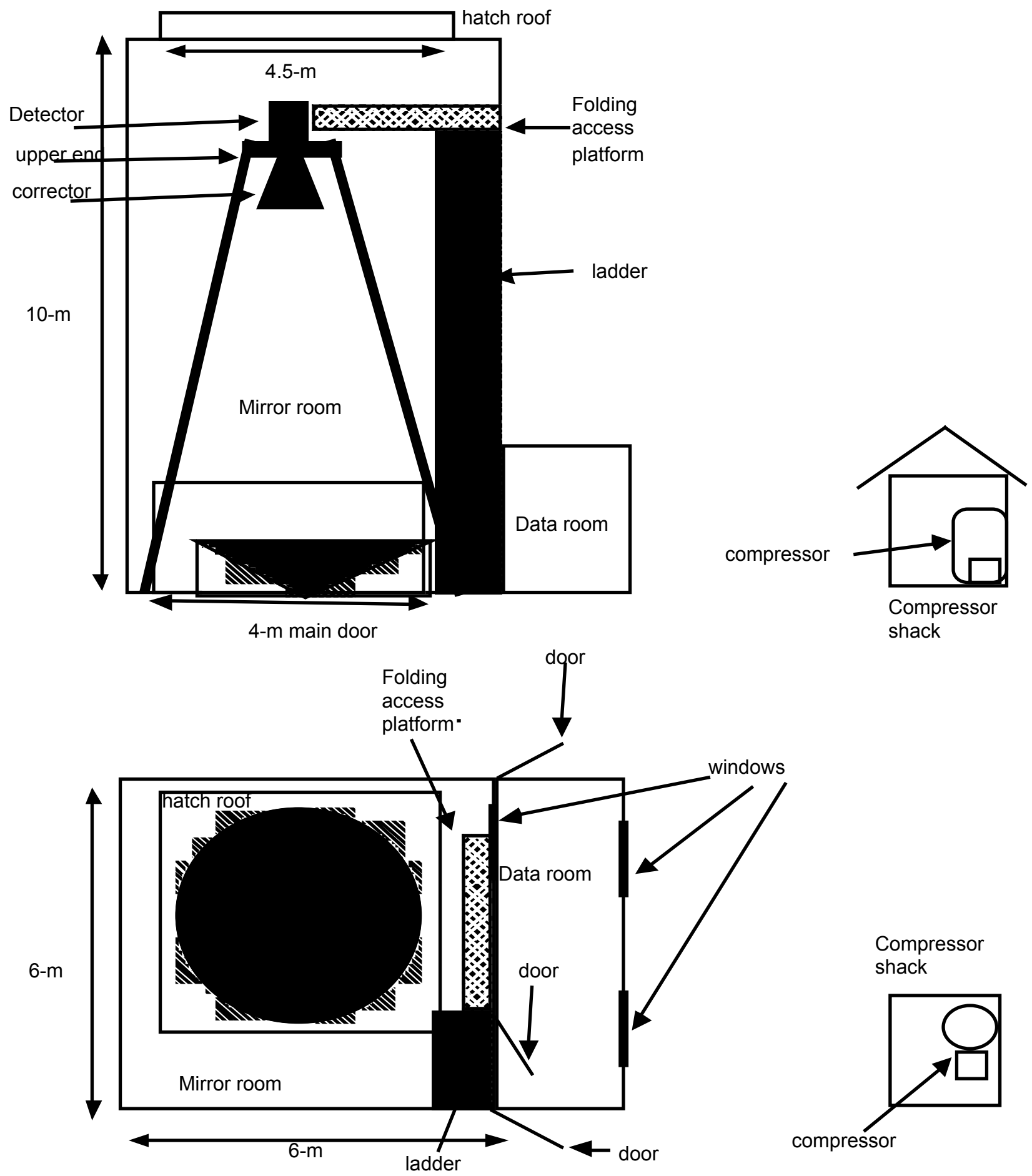

Figure AI.3: It shows a layout of the telescope and observatory for a 4-m telescope.

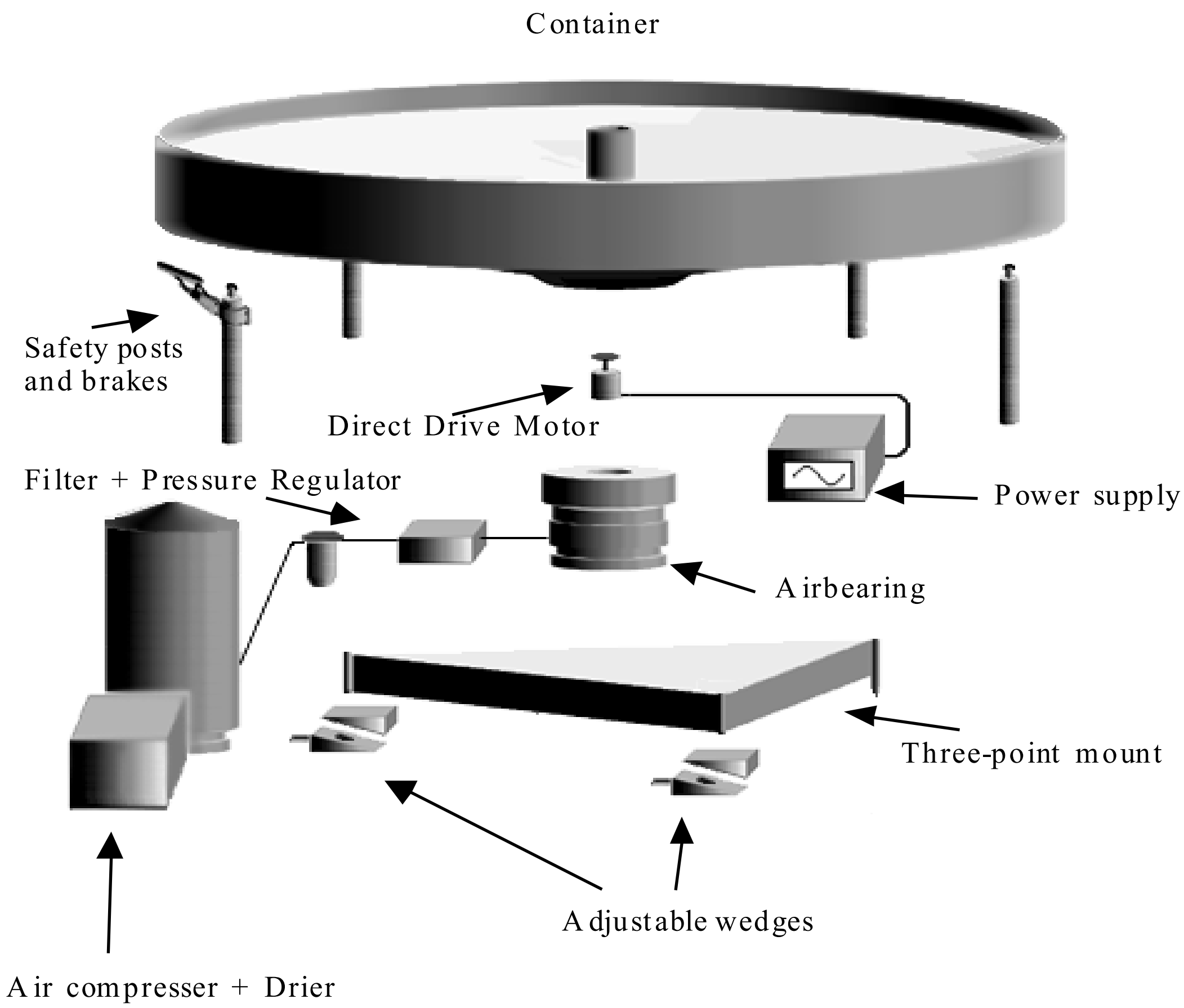

Figure AI3. It shows an exploded view of a basic liquid mirror setup.

**AI.1 Procedures used to estimate the costs in table 4.**

All costs are in US$. They do not include shipping costs.

**Building:**The building is a prefabricated single wall steel structure. The estimated cost includes installation costs. To arrive at an estimate, I discussed the problem at length with an architect. Following his advice, I contacted a local firm, discussed the building with them and asked for a quote for the assembled building. Assembly costs have been calculated assuming the costs of Canadian manpower and translated to US dollars. The estimate seems reasonable when compared to with the costs incurred for the enclosures of the smaller observatories described in appendix II.

**Mirror:** Costs are based on the materials and labor hours used to build the 3.7-m in the Laval liquid mirror laboratory (Tremblay 1999). About half the cost comes from labor, assuming US$45/hour. Labor costs can be significantly lower since most labor goes to building the container, a simple task that can be done with unskilled low-wages labor (e.g. graduate and summer students). Graduate student R. Tremblay (Tremblay 1999) built most of the container of the 3.7-m. There is some uncertainty in the cost of the airbearing. The estimate assumes the worse case scenario that we must use a large airbearing made by Professional Instruments Co. that costs 65,000 US $. This bearing is way too strong for a 4-m. It is possible to have a smaller, less expensive airbearing bearing custom designed. It is not necessary to design the composite material containers for mirrors smaller than 6-m since they are given by Content (2003) and Borra et al. (2003). Tremblay (1999) gives a detailed description of how he built his container,

**Upper-end:** I called the company that built the upper-end of the Canada-France-Hawaii 3.6-m telescope and asked for an estimate. This is certainly an overestimate since, unlike the CFHT upper end, the upper end of a zenith telescope does not swing around and can thus be much simpler. In particular, it can use commercial optical mounts; as done by NODO.

**Frame**: It is a such a simple structure. The guesstimate of 10,000 $ seems reasonable.

**Corrector:** The cost of a corrector is somewhat uncertain since it is a custom made item. Quotes on the same element can vary considerably from a supplier to another. Things are further complicated by the fact that the corrector of a zenith telescope is unusual since it must also include the TDI correction (Hickson & Richardson 1998) namely must introduce just the right amount of distortion. To get a costing estimate one should design the corrector and ask for a quote. To come up with a realistic estimate, I have consulted with Drs. Paul Hickson and Harvey Richardson, who is arguably one of the most respected optical designer of astronomical optics. He has designed correctors for a 5.1-m and a 6-m zenith telescopes. A quote for 60 USK$ for the polished glass was secured from a Canadian company (APS). On the bases of that estimate and the design of the corrector made for a 4-m by Richardson, I estimate that the corrector should cost about 100,000 US$).

**Miscellanea:** This includes things like shipping costs, filters, contingency, etc….

### AI.2 Comments:

In private conversations, many have expressed skepticism as to the "real costs" of building and operating (see Appendix II) an LMT observatory in the "real world". Such skepticism is understandable for the costs quoted are so much smaller than the costs associated with running a classical telescope. An indication of the "real costs" of building and operating a 4-m class LMT can be gathered from the NODO experience. The total cost to NASA of building and operating that telescope for seven years was 2 million US dollars (Mulrooney, private communication). That included construction and a significant amount of research and development on this first generation LMT. For example, they experimented with different bearings, designed and build a novel type of direct drive. They built an observatory near the NASA headquarters and debugged it there before moving it to its final site. The cost also included the salary of the research scientist for a few years. Costs can obviously be substantially lower for a University run observatory. It must be noted that NODO was built and operated by NASA standards. In particular, they applied NASA safety standards insofar as mercury vapors are concerned (Mulrooney 2000). A university type of construction and operation can be significantly less expensive.

## APPENDIX II. Operating Costs of a 4-m Observatory Using a 4-m Liquid Mirror Zenith Telescope

We can make a reasonable estimate of operating costs because 4 LMT observatories have been operated for several years. Two of them are used for lidar research in the atmospheric sciences (Sica et al. 1995, Wuerker 1997). Two of them (UBC-LAVAL 2.65-m, Hickson et al. 1994, and NASA Optical Debris Observatory (NODO) , Hickson & Mulrooney 1997) have been used for astronomical research. The NODO 3-m is particularly relevant because its operation, although it underwent significant R&D at its beginnings, was a professionally run operation. Table 5 was drawn from a budget suggested by Dr. M. Mulrooney, based on his extensive experience with the operation of NODO. I add below some discussion of the past experience with LMT observatories and of what is involved in operating them.

### II.1. Previous Experience with LMT Observatories.

**UCLA and UWO Lidars**: Although they do not obtain images the experiences with the UCLA and UWO 2.6-m are relevant to this article. They are happy with their mirrors and say that they are robust and have low maintenance costs.

**UBC-Laval 2.65-m**. The telescope has operated for two observing seasons. It was an experimental system since it was the first astronomical LMT. The observatory was near Vancouver, Canada. It was operated by a graduate student (Cabanac, 1998) and P. Hickson.

**NASA NODO 3-m.** This is by far the most relevant LMT because it has been operated for 7 years as an imager. Besides data gathered to observe space debris, it has taken massive quantities of astronomical images a substantial fraction of which has been analyzed and published. It is located on an isolated mountain top in New Mexico but other astronomical facilities are located 20 miles away. The 3-m LMT is a first generation LMT and much of it was experimental. For a few years graduate student M. Mulrooney (2000) operated it (including debugging) with minimal help. He took care of the mirror, the CCD, the telescope itself, etc…During the last years of operation they used a telescope operator but there is no reason why the telescope should not run automatically or semi automatically. A mature system (such as NODO in its last years) requires as little as 3hours/week maintenance (Mulrooney, private communication).

**II.2 Maintenance.**

Figures AI.1, AI.2 and AI.3 show schemqtics of mirror, enclosure and telescope, illustrating its simplicity. The telescope frame is a vertical tower, needing essentially no maintenance. We can expect occasional adjustments to the optical elements of the corrector and occasional maintenance to the focusing system. The detector and data taking hardware and software are the usual ones, with the usual associated operating and maintenance costs. There is no instrumentation change (only occasional filter changes), thus simplifying operation. The cooling unit can consist of a commercial laboratory unit positioned close to the detector, resulting in lower costs and maintenance. The enclosure has few moving parts (asliding roof and a small movable platform). Its maintenance is therefore significantly lower than for a rotating dome.

Liquid mirrors have been operated for years in observatory settings (including cold Canadian winters by Sica et al. 1995) so that we accurately know their performance and their maintenance costs. Costs are very low. The mirror must be cleaned once a month (requiring one person and about an hour ). There is some maintenance, and the occasional problems, involved with any mechanical system (e.g. the compressor). But they do not take much time cost very little since the mechanical systems are simple. All users of liquid mirrors are unanimous in their assessment that the mirror is the most reliable and less troublesome part of their operations.

**II.3. Operation.**

An operator is needed to open the hatch sufficiently early that the heat of day will have dissipated when observing starts. The operator must then start observing and occasionally monitor the operation. He must close at the end of the night. If the data acquisition system uses tapes for data storage, he must load and download them. About once a month he must stop the mirror, clean it and restart it. This takes about one hour. There must be regular maintenance of the CCD, the data acquisition system as well as the building. In a site where there are other telescopes, the observatory could be run part time by the operator of an other telescope.

**II. 3 Comments**

One can get an appreciation of the resources needed for the operation of an LMT observatory by noting how NODO was run during the last few years. The project scientist lived 1300 km away and spent about 1 week a month on the mountain to do required maintenance and some R&D. He took care of maintenance to the entire system (LM, CCD, etc…) during his week stay on the mountain. There was a single night observer on the mountain and, essentially, her only activity consisted in taking data.

Table 1

Broad-band limiting magnitudes.

(4-m telescope with a 4096x4096 CCD mosaic having 0.4-arcsecond pixels, S/N = 5)

| 1 night (127 sec) * | | | 1 season (7600 sec)* | | | 4 seasons (30,500 sec)* | | |
|---|---|---|---|---|---|---|---|---|
| B | R | I | B | R | I | B | R | I |
| 24.4 | 24.1 | 23.7 | 26.6 | 26.3 | 25.9 | 27.4 | 27.1 | 26.7 |

* Because observations in I can be carried out during moonlit nights, I have assumed integration times twice as long in I.

Table 2

Apparent R and I magnitude of a type Ia supernova at maximum light as function of redshift

| z | $m_R$ | $m_I$ |
|---|---|---|
| 1.0 | 25 | 23.8 |
| 0.8 | 23.8 | 23.0 |
| 0.6 | 22.7 | 22.5 |
| 0.4 | 21.8 | 21.6 |
| 0.2 | 20.2 | 20.2 |

Table 3

Predicted SNe Ia discovery rates in $\Delta z = 0.2$ bins.

| z | events/year |
| --- | --- |
| 0.2 | 500 |
| 0.4 | 1000 |
| 0.6 | 1500 |
| 0.8 | 1900 |
| 1.0 | 2000 |

Table 4
Cost breakdown of a 4-m telescope and observatory
(Costs are in units of 1000.0 US$)

| Item | Cost | Based on* |
|---|---|---|
| Building: | 75 | Design and quotes |
| Mirror: | 110 | U. Laval 3.7-m |
| Upper end: | 110 | CFHT upper end |
| Corrector: | 100 | Design and quote |
| Frame: | 10 | guesstimate |
| Miscellanea | 50 | guesstimate |
| Grand Total | 555 | |

*Appendix I discusses how these estimates were obtained. The detector and the data acquisition system are not included.

TABLE 5

Estimated Yearly Operating costs of a Zenith Telescope Using a 4-m Liquid Mirror
(Costs are in units of 1000.0 US$)

| LABOR | | UTILITIES | | Equipment+Consumable * | |
|---|---|---|---|---|---|
| Site Maintenance | 9.4 | Electric | 1.5 | Exabyte tapes | 1.8 |
| Site Engineering | 7.4 | Telecom | 2.4 | Safety | 1.0 |
| Machinist (Skilled) | 7.8 | Propane | 2.0 | Miscel. Cons. | 0.5 |
| Operator | ??? | Overhead | 1.2 | Shipping | 5.0 |
| | | | | Overhead | 1.7 |
| Total | 24.6+? | Total | 7.1 | Total | 10 |
| Grand Total | 41.7 + Operator(s) salaries + Travel costs | | | | |

*Appendix II discusses how these estimates were obtained

Figure 1

It converts equatorial into galactic coordinates and can be used to determine the regions of sky sampled by a zenith telescope in a given site. As the Earth rotates and the seasons change, the telescope scans a strip of constant declination moving in and out of the galactic plane.

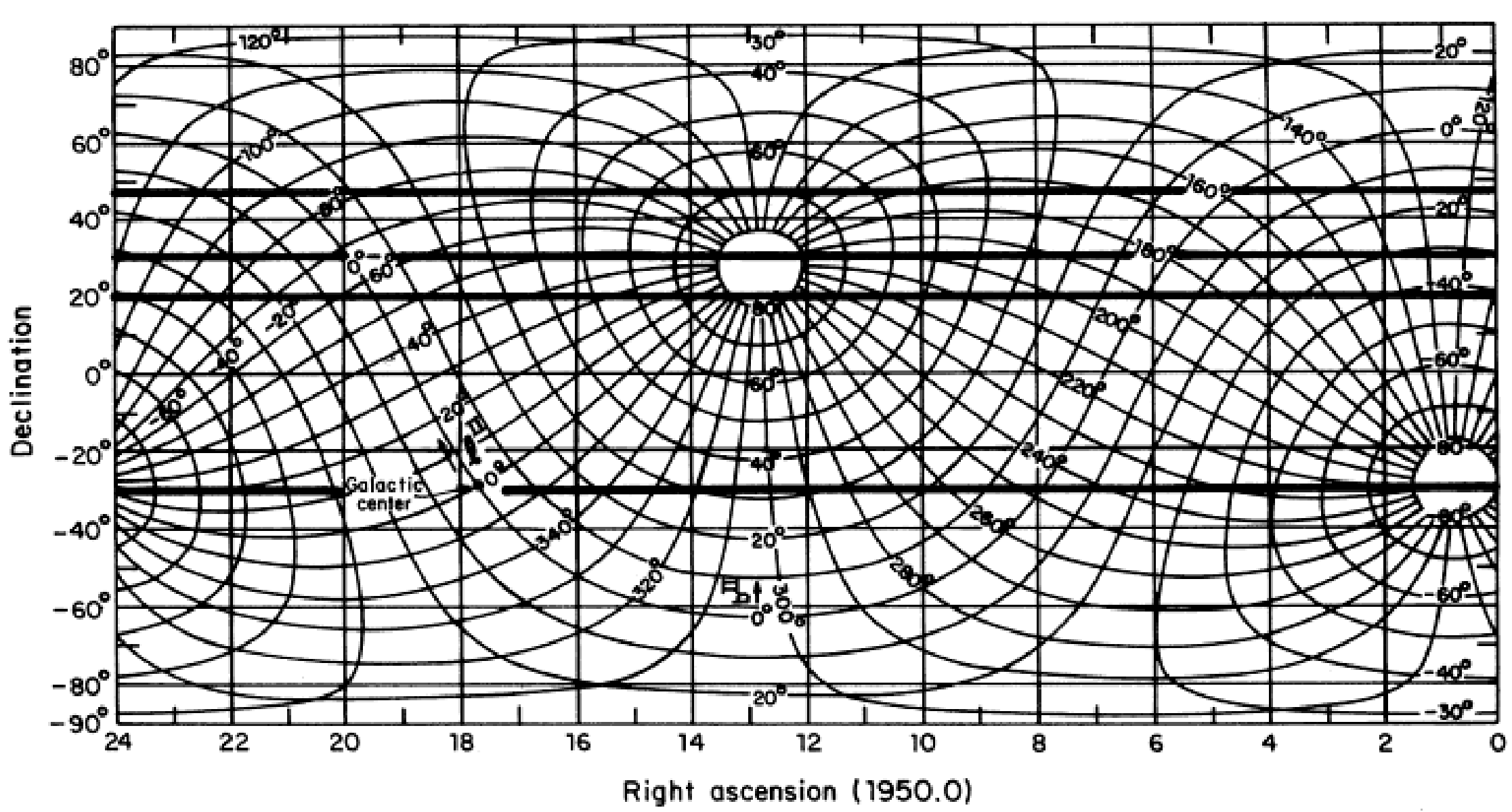

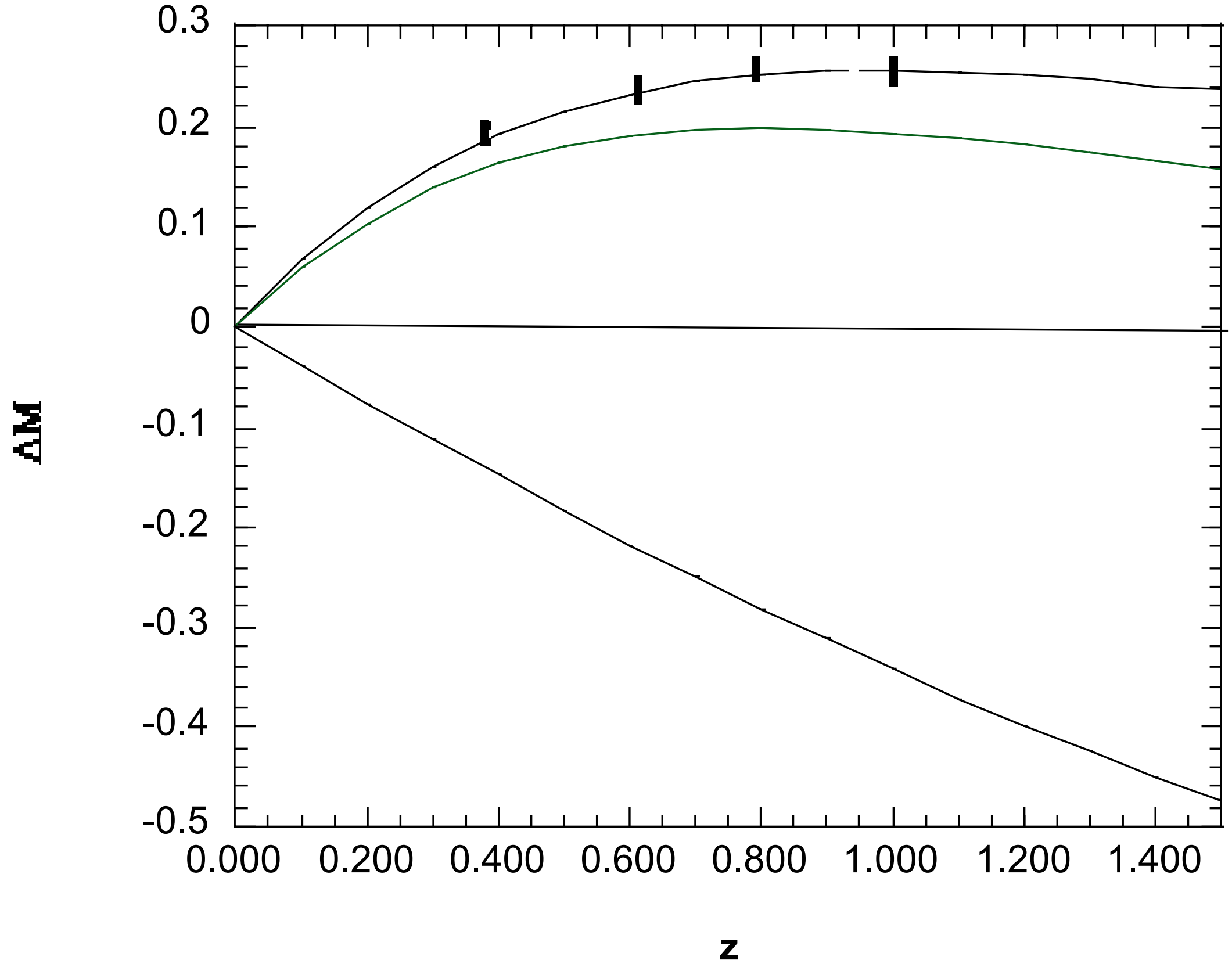

Figure 2

It gives the magnitude-redshift relations for three “flat” cosmological models with different combinations of $\Omega_M$ and $\Omega_\Lambda$ (top to bottom $\Omega_M$ =0.3, $\Omega_\Lambda$= 0.7; $\Omega_M$ = 0.35, $\Omega_\Lambda$=0.65, $\Omega_M$ =1.0, $\Omega_\Lambda$=0.0). For clarity, the magnitude-redshift relation for the $\Omega_M$ = 0.3, $\Omega_\Lambda$= 0.0 model has been subtracted from all curves. The small vertical bars give 2 standard deviations error bars estimated for the supernova survey discussed in section 4.

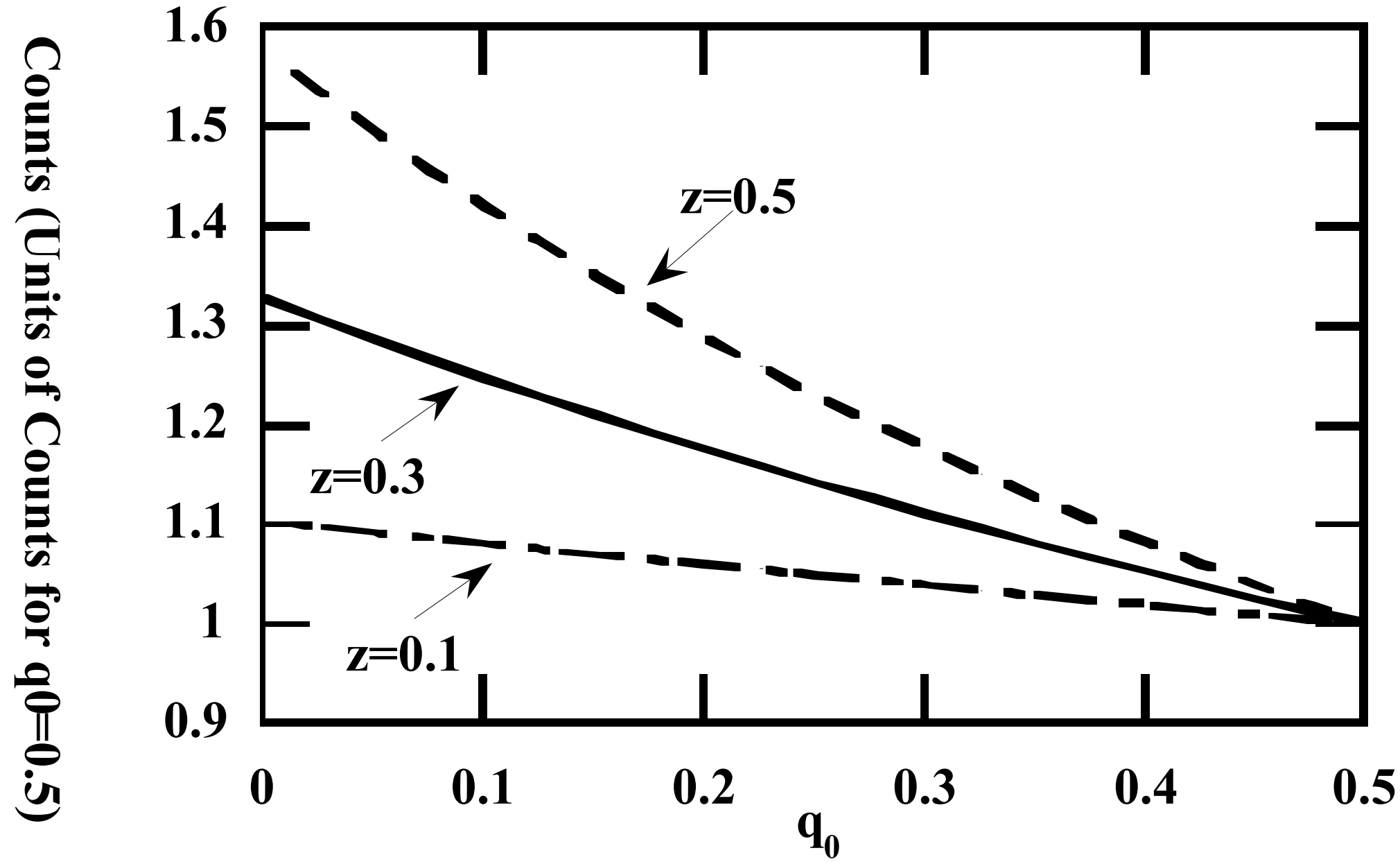

Figure 3

The 3 curves show the galaxy counts predicted at z = 0.1, 0.3 and 0.5 as functions of $q_0$, normalized to the counts predicted for $q_0$=0.5.

Figure 4

Redshift distributions expected for surveys reaching 22nd, 24th and 27th blue magnitudes.

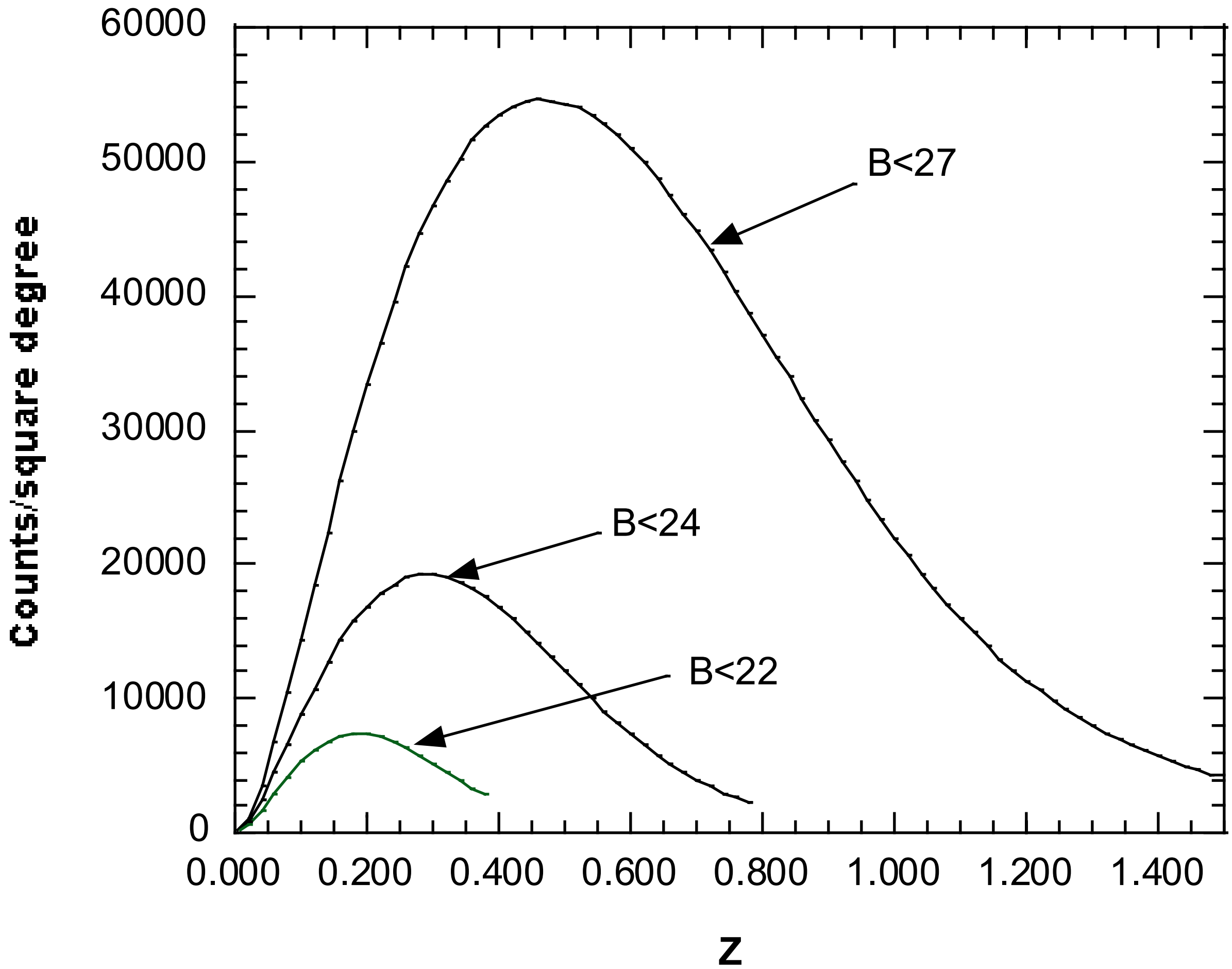

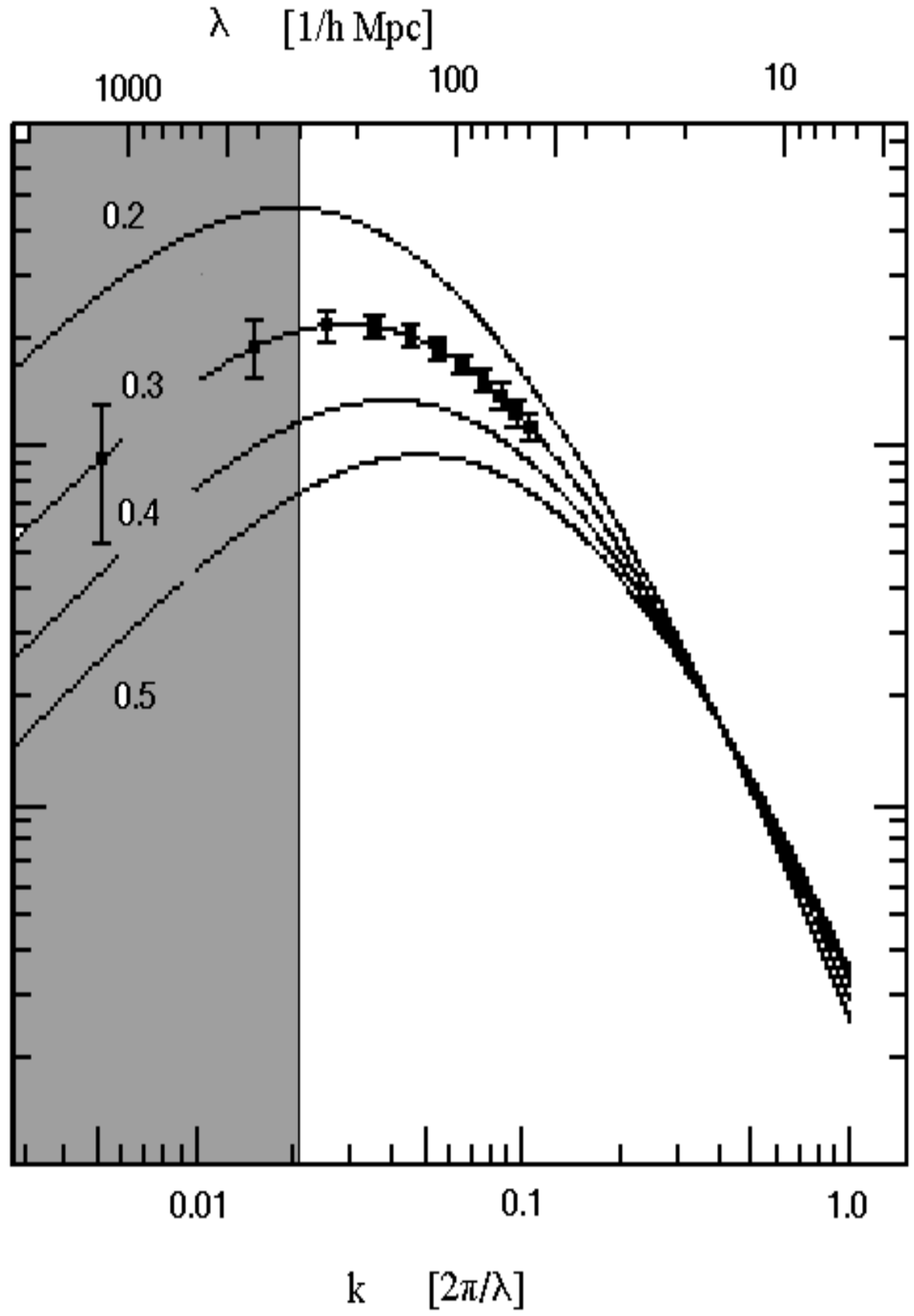

Figure 5: It compare 1 σ uncertainties, expected for a volume limited (to M*) sample of the SDSS northern redshift survey and assuming Gaussian fluctuations, to power spectra for CDM with different Ωh. The shaded box shows the range of the HPBW of the z distribution of the LZT survey. Because the total numbers of galaxies are similar in the 2 surveys, we can expect a similar distribution of error bars shifted to the shaded box.